\documentclass[12pt,preprint]{aastex}

\slugcomment{submitted to ApJ}

\shorttitle{Pulsation and Evolutionary Masses of Cepheids}
\shortauthors{Keller}

\begin{document}


\title{Cepheid Mass-loss and the Pulsation -- Evolutionary Mass Discrepancy}


\author{Stefan C.\ Keller}
\affil{RSAA, Australian National University, Canberra A.C.T.~2600, Australia}



\begin{abstract}
I investigate the discrepancy between the evolution and pulsation masses for Cepheid variables. A number of recent works have proposed that non-canonical mass-loss can account for the mass discrepancy. This mass-loss would be such that a 5$M_{\odot}$ star loses approximately 20\% of its mass by arriving at the Cepheid instability strip; a 14$M_{\odot}$ star, none. Such findings would pose a serious challenge to our understanding of mass-loss. I revisit these results in light of the Padova stellar evolutionary models and find evolutionary masses are ($17\pm5$)\% greater than pulsation masses for Cepheids between $5<M/M_{\odot}<14$. I find that mild internal mixing in the main-sequence progenitor of the Cepheid are able to account for this mass discrepancy.

\end{abstract}




\section{Introduction}

$\delta$ Cepheid variables are an essential step in our determination of extragalactic distances. Apart from their use as distance indicators, the regularity of Cepheid pulsation provides a well-defined set of observational parameters with which to probe the course of stellar evolution for intermediate mass stars.

Over the preceeding decades much effort has been devoted to reconciling mass determinations for Cepheids from the various methods at our disposal (see \citet{Cox80} for a review). The longest standing of these, the Cepheid pulsation mass discrepancy was first revealed by \cite{Stobie69} who showed that pulsation masses for bump Cepheids\footnote{Bump Cepheids are distinguished by a secondary local maximum of the light curve seen in Cepheids with periods in the range of 6 to 16 days \citep{Bono00b,Hertzsprung26}.} were significantly lower than those predicted by stellar evolutionary models. \cite{Andreasen88} showed that by artifically enhancing the opacity over the temperature range $1.5\times10^{5}K<T<8\times10^{5}K$ by a factor of 2.5 it was possible to remove the discrepancy. More detailed modelling of opacities by the OPAL \citep{Rogers92} and the Opacity Project \citep{Seaton94} confirmed an increase in this temperature range due to metal
opacity. 

The implementation of new opacities largely resolved the bump Cepheid mass discrepancy \citep{Moskalik92}. However, despite convergence, a number of subsequent studies have shown that the discrepancy remains significant and requires explanation. Studies of Galactic \citep{Natale07,Caputo05}, LMC \citep{Wood97,Keller02,Keller06,Bono02} and SMC \citep{Keller06} Cepheids have shown that the masses determined via pulsation modelling are $\sim$15-20\% less massive than those expected from evolutionary models. Dynamical masses for Cepheids are difficult to obtain given the low spatial density of Cepheids. The works of \citet{Benedict07,Evans07, Evans06, Evans98} present dynamical masses for six Cepheids. Albeit with large associated uncertainties, these results confirm the conclusions drawn for pulsation modelling; that the evolutionary masses appear $\sim15$\% larger. 

From an evolutionary perspective, Cepheids are understood to be post-red giant stars crossing the instability strip on so-called blue loops following the initiation of core-He burning. To ascribe an evolutionary mass one takes the Cepheid's luminosity and, using a mass-luminosity (M-L) relation that is derived from evolutionary models, derives the mass of the Cepheid. The M-L relation can be modified substantially by the treatment of internal mixing and mass-loss. Both processes feature complex hydrodynamical and radiative mechanisms for which we, at present, only possess empirical approximations.

The treatment of internal mixing modifies the size of the helium core established during the star's main-sequence (MS) evolution. Overshoot at the edge of the convective core of the Cepheid progenitor mixes additional hydrogen into the core and hence increases the helium core mass. As a consequence the post-MS evolution occurs at a higher luminosity. Mass loss, in an ad-hoc manner at least, offers a mechanism to modify the M-L relation by directly reducing the mass of a Cepheid. 

The properties of pulsation, on the other hand, are dependent on the structure of the atmosphere of the Cepheid. \citet{Keller06} show that the morphology of a bump Cepheid light curve is highly sensitive to the mass, luminosity, effective temperature and metallicity. Hence modelling of the light curve can be used to determine a pulsation Cepheid mass that is entirely independent of stellar evolution calculations. 

In the work of \cite{Bono02} and \cite{Caputo05} it is proposed that mass-loss can account for the mass discrepancy between pulsation and evolutionary masses. \citet{Caputo05} also conclude that models that incorporate additional internal mixing in the vicinity of the convective core are not able to explain the mass discrepancy. In this paper we revisit these conclusions and present a scenario for resolution of the mass discrepancy.

\section{The Cepheid Pulsation -- Evolutionary Mass Discrepancy}

\begin{figure}
\begin{center}
\includegraphics[scale=0.80, angle=0]{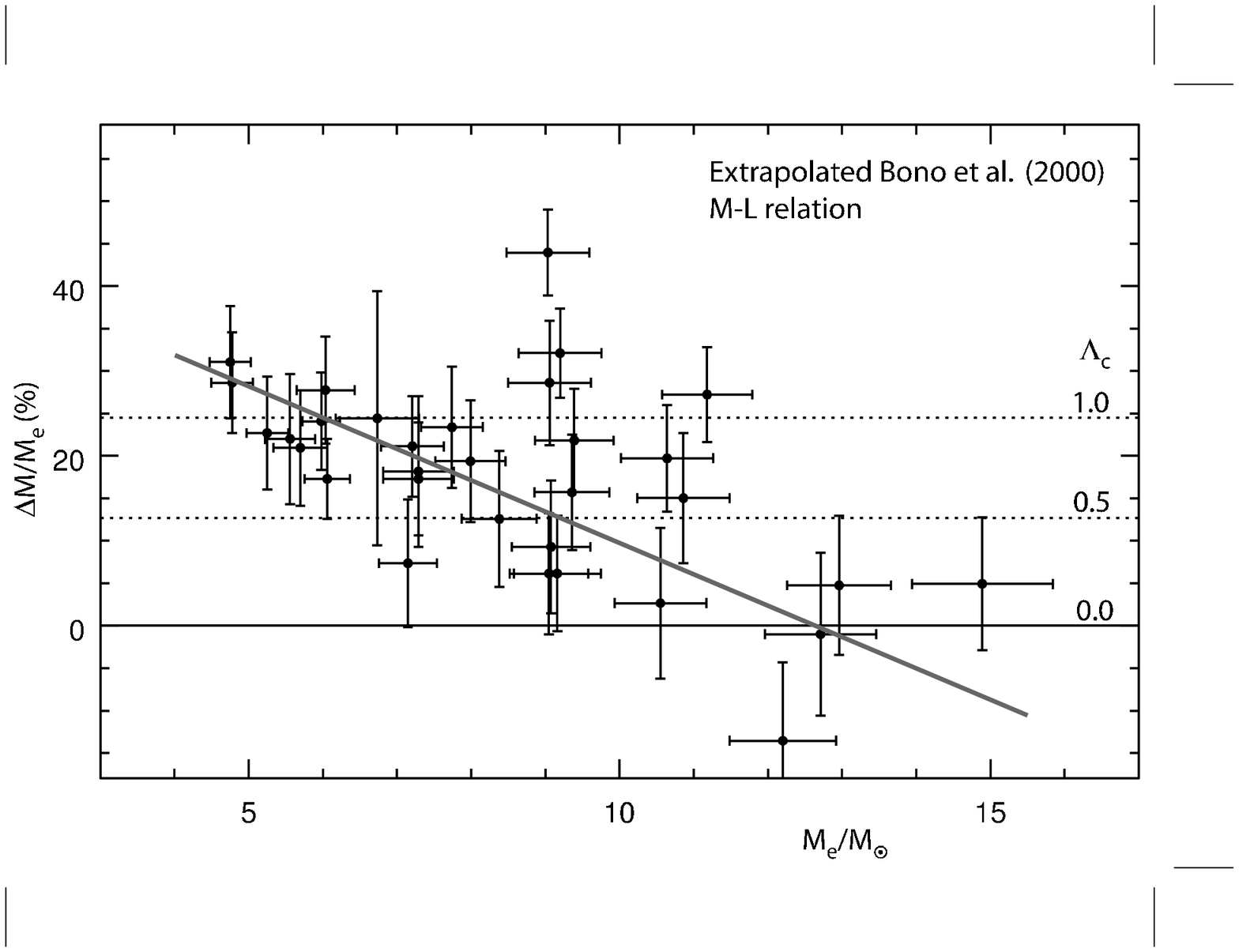}
\caption{Here we show the mass discrepancy as a function of mass. $\Delta M/M_{E}$ expresses the mass discrepancy and is equivalent to the difference between the pulsation mass, $M_{P}$, and classical evolutionary mass, $M_{E}$ (i.e.\ $M_{E}$ does not incorporate convective core overshoot), normalised by $M_{E}$. Here the extrapolated mass-luminosity relation of \citet{Bono00a} was used (the dashed line in Figure \ref{figure:ML}, see text for details). Note the mass discrepancy vanishes at higher masses. Overlaid are the locii of models that incorporate mild ($\Lambda=0.5$; \citet{Girardi00}) and moderate ($\Lambda=1.0$; \citet{Bressan01}) convective core overshoot.}
\label{figure:Caputo}
\end{center}
\end{figure}

A useful way of expressing the mass discrepancy is to form the quantity $\Delta M/M_{E}$ in which $\Delta M$ is the difference between the pulsation mass, $M_{P}$, and the canonical (i.e. not including CCO) evolutionary mass, $M_{E}$, as shown in Figure \ref{figure:Caputo}. In Figure \ref{figure:Caputo} we also show the effects of the inclusion of mild ($\Lambda_{c}$=0.5) and moderate ($\Lambda_{c}$=1.0) CCO. The effect of CCO is to raise the luminosity of the Cepheid of a given mass as discussed above. 

\citet{Caputo05} utilise the $BVIJK$ absolute magnitudes derived by \cite{Storm04} using distances from the near-infrared surface brightness (IRSB) method. \cite{Caputo05} have used the mass dependent period-luminosity-color relation to determine the pulsational Cepheid mass using the M-L relation of \citet{Bono00a}. They find that $\Delta M/M_{E}$ ranges from 20\% at $M\sim5M_{\odot}$ to $\sim$0\% at $M\sim14M_{\odot}$ (see Figure \ref{figure:Caputo}). Such a trend can not be explained by a uniform level of core overshoot. Monte-Carlo simulation using the individual quoted uncertainties of the data shown in Figure \ref{figure:Caputo} reveals a gradient of $(3.4\pm0.5)$\%$M_{\odot}^{-1}$. 

A caveat to IRSB analysis that underlies the results of \citet{Storm04} and \citet{Caputo05} is the necessary introduction of  the poorly understood projection factor, $p$, that embodies the effects of limb-darkening \citep{Nardetto06}, and is dependent on the pulsation velocity and the species under study \citep{Nardetto07}. $p$ is approximated as a function of period, and it has been speculated that this could introduce a period dependency in the derived distances. However, the work of \cite{Fouque07} shows that direct HST parallaxes for a sample of Cepheids agree within uncertainties with the IRSB measures albeit for only the five stars with distance determinations from both techniques.

To explain the mass discrepancy found at lower masses, \citet{Caputo05}, propose mass-loss from the Cepheid progenitor  before, or during the central He-burning phase. The implied total mass-loss {\emph{declines}} from $\sim20$\% at $M\sim5M_{\odot}$ and vanishes by $M\sim14M_{\odot}$. Such mass-loss is seemingly at odds with empirical estimates of mass-loss rates which show that mass-loss {\emph{increases}} with stellar luminosity and radius \citep{Reimers75, deJager88, Schroeder07}. Furthermore, Caputo et al. conclude that CCO can not account for this trend in $\Delta M/M_{E}$ with mass since this would lead to unphysical $\Delta M/M_{E,\Lambda} < 0$ (their Figure 9; where $\Delta M/M_{E,\Lambda}$ is the mass at a given $\Lambda_{c}$) for higher mass Cepheids. 

\begin{figure}
\begin{center}
\includegraphics[scale=0.60, angle=0]{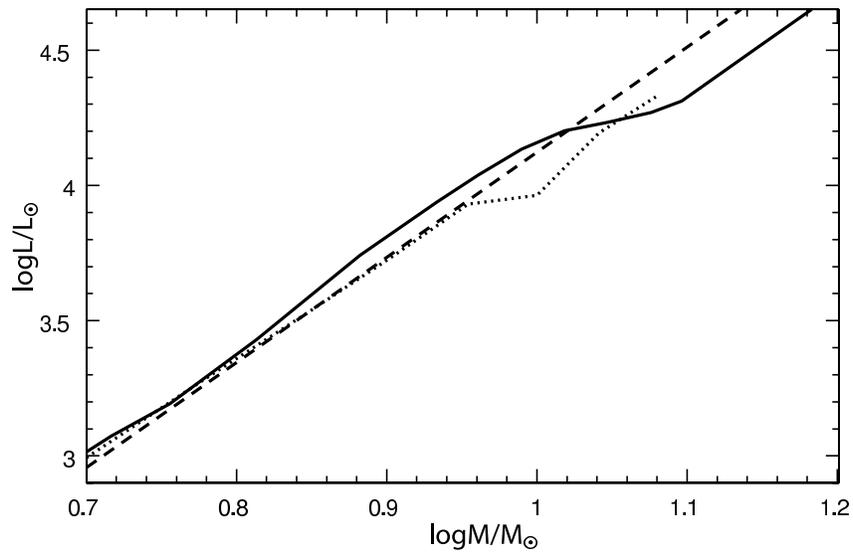}
\caption{The Cepheid mass-luminosity relation used by \citet{Caputo05}
  consists of the dashed line (from Caputo et al.\ Equation 2). The models of
  \citet[$Z$=0.02, $Y$=0.27 and $\Lambda_{c}=0$]{Bono00a} without mass-loss are shown by the dotted line. The \citet{Bressan93} mass-luminosity relationship for Cepheids is shown as the solid line. Note the significant departures from the linear relation of relation of Caputo et al. }\label{figure:ML}
\end{center}
\end{figure}

The Cepheid M-L relation implemented in analysis is critical. In their study, \cite{Caputo05} have chosen to utilise a linear M-L relation derived from \citet{Bono00a} evolutionary models ($Z$=0.02, $Y$=0.28 and $\Lambda_{c}$=0) that do not incorporate mass loss. This M-L relation is shown as the dashed line in Figure \ref{figure:ML}. However, as shown by the evolutionary models of \cite{Bono00a} there are significant departures from this linear relationship for $M \ge 9M_{\odot}$ (see the dotted line in Figure \ref{figure:ML}). 

In the analysis to follow we incorporate the non-linear nature of the Cepheid
M-L into our analysis. The models of \cite[$Z$=0.02, $Y$=0.27 and $\Lambda_{c}=0$]{Bono00a} extend to $M$=12$M_{\odot}$, however the derived pulsation masses reach to 14.9$M_{\odot}$. Let us consider the stellar evolutionary sequences of \citet{Bressan93} that extend to higher masses. These models ($Z$=0.02, $Y$=0.28 and $\Lambda_{c}$=0) implement mass-loss according to the \citet{Reimers75} and \citet{deJager88} formulation and are shown in Figure \ref{figure:ML} as the solid line. For masses less than 8$M_{\odot}$ there is clearly a different gradient between the \cite{Bono00a} and \cite{Bressan93} M-L, a feature discussed by \cite{Beaulieu01}. What is common to both models is a marked departure from a linear M-L in the mass range 9-10$M_{\odot}$. This departure results in evolutionary masses that are greater for a given luminosity compared with those derived from the linear M-L for masses greater than 9-10$M_{\odot}$.

\begin{figure}
\begin{center}
\includegraphics[scale=0.80, angle=0]{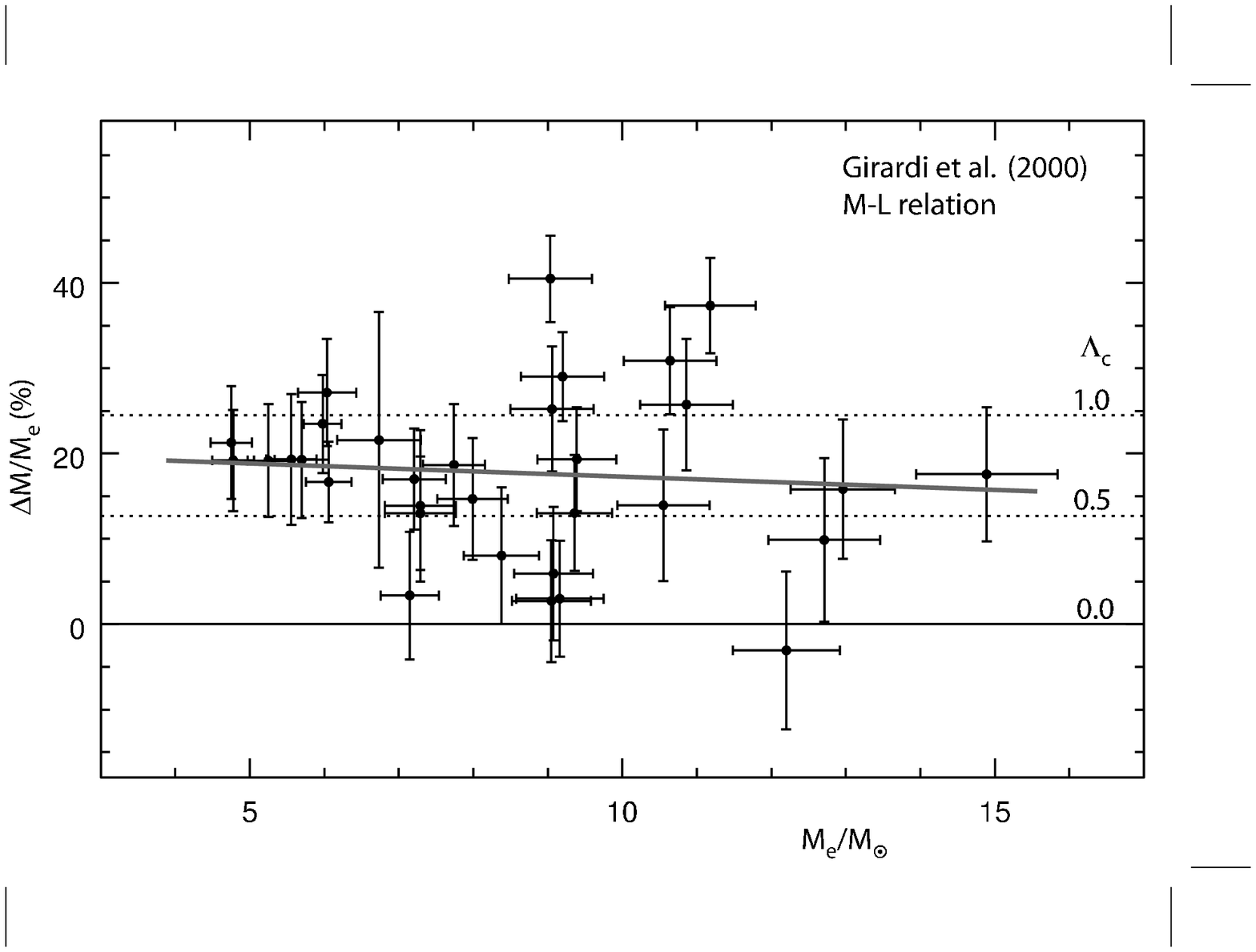}
\caption{As in Figure \ref{figure:Caputo} except here the mass-luminosity relation is due to \citet{Bressan93}. The mass discrepancy shows no significant dependence on mass, but rather a uniform offset of ($17\pm5$)\% (see text for details).}
\label{figure:Girardi}
\end{center}
\end{figure}

The use of the \citet{Bressan93} M-L results in Figure \ref{figure:Girardi}.  Figure \ref{figure:Girardi} reveals no significant trend in $\Delta M/M_{E}$ as a function of $M_{E}$ (Monte-Carlo simulation of the data and associated uncertainties shown in Figure \ref{figure:Girardi} reveals a gradient of $(0.3\pm0.5)$\%$M_{\odot}^{-1}$). This demonstrates that a consideration of the non-linear Cepheid M-L accounts for the decrease in $\Delta M/M_{E}$ seen in the work of Caputo et al. The absence of a significant gradient also negates the argument by \citet{Caputo05} that CCO can not be the source of the Cepheid mass discrepancy. 

Reanalysis of the results of \citet{Caputo05} shows that the evolutionary Cepheid mass is ($17\pm5$)\% greater than that predicted from pulsation modelling. In the next section I discuss the merits of a range of possible causes.

\section{The Source of the Cepheid Pulsation -- Evolutionary Mass Discrepancy}

The hiding place of the source of discrepancy between pulsation and evolutionary Cepheid masses has shrunk dramatically over the last three decades. To account for the discrepancy we have three key options. Firstly, it is possible that some input physics in pulsation calculations are not sufficiently described. In this regard, the only input that could affect pulsation to this magnitude is the description of radiative opacity. Second, that the Cepheid mass is smaller than their main-sequence progenitors due to mass-loss. And finally, that evolutionary calculations underestimate the mass of the He core in intermediate mass stars and so underestimate their luminosity. We now discuss each of these possible sources in detail.

\subsection{Radiative Opacity}

The pulsation properties of Cepheids are critically dependent on the so-called $Z$-bump opacity arising from the dense spectrum of transitions originating from highly ionized Fe. The inclusion of these transitions in the works of OPAL \citep{Rogers92} and OP \citep{Seaton94} resulted in a substantial increase in opacity at log$T\approx5.2$. The Opacity Project \citep{Badnell05} has included further details of atomic structure (in particular, the treatment of atomic inner shell processes) in their calculation of opacity. The new opacities do show an increase over the 1992 OP and OPAL values of opacity in the $Z$-bump, but at a level of only 5-10\% \citep{Badnell05}. To account for the mass discrepancy the opacity would need to be raised by 40-50\%, equivalent to the increase between the early Los Alamos opacities \citep{Cox76} and OP and OPAL opacities. Hence the uncertainty in radiative opacity is an unlikely resolution to the Cepheid mass discrepancy.

\subsection{Mass-Loss}

The studies of \citet{Bono02, Bono06} and \citet{Caputo05} propose that mass-loss can account for the mass discrepancy. Candidate mass-loss phases include the red giant branch phase, subsequent blue loop evolution, or possibly from the action of pulsation itself \citep{Bono06}. The removal of mass from the Cepheid is a straightforward, albeit ad-hoc, way to bring the evolutionary mass in line with that derived from pulsation. The timing of, and the changes in stellar structure brought about by significant mass-loss are important to the net change in the Cepheid M-L. For instance, significant mass-loss on the MS causes a reduction in overall mass and hence helium core mass, resulting in a reduction in luminosity in the instability strip \citep{de-Loore88}. Enhanced mass-loss on the red giant branch reduces envelop mass, and the material available to the hydrogen-burning shell within the Cepheid, again leading to a reduction in luminosity relative to a star without mass-loss \citep{Yong00}. One of the difficulties with the proposal for mass-loss to solve the Cepheid mass discrepancy is that it would require the rather artificial bulk removal of material without consequent modifications to stellar structure and energy production.

Furthermore, standard mass-loss can account for at most a few percent reduction in Cepheid mass and not the 15\%-20\% required. Mass loss is usually treated the semi-empirical relation of \citet[`Reimers' law']{Reimers75}. While not providing any physical reasoning on why the mass-loss is generated, `Reimers' law' provides an adequate match to observed mass-loss rates over a broad range of stellar parameters \citep{Schroeder07}. 

Major mass-loss is expected during the red giant branch (RGB) evolution. The models
of \cite{Girardi00} and \cite{Bressan93} use a parameterised, empirical fit $dM/dt = -4\times10^{-13}\eta L/gR$ \citep{Reimers75}. The value of $\eta$ is set by a consideration of the masses of stars on the horizontal branch (HB) of globular clusters. Determination of $\eta$ is made difficult due to the variety of HB morphology exhibited by globular clusters. If we consider the distribution of effective temperatures for HB stars is entirely due to variable mass-loss then $\eta$ must range from 0 to somewhat more than 0.4.  Using the canonical value of $\eta=0.4$, a 5$M_{\odot}$ star looses $\sim$0.03$M_{\odot}$ during the RGB phase. To accomodate the mass discrepancy seen in Figure \ref{figure:Caputo} this would have to be increased to 0.8$M_{\odot}$ corresponding to a 20-30 fold increase in $\eta$ which is not plausible. Therefore mass-loss on the RGB does not resolve the mass discrepancy.

\citet{Caputo05} and \citet{Bono06} suggest that pulsation may give rise to an enhancement of mass-loss. Attempts to measure the mass-loss rates for Cepheids have been made using IRAS infrared excesses \citep{McAlary86} and IUE UV line profiles \citep{Deasy88} and in the radio \citep{Welch88}. \cite{Deasy88} found mass-loss rates for the majority of Cepheids of the order of a few $\times$ 10$^{-9}$ $M_{\odot}$yr$^{-1}$. The study of \cite{Welch88} places upper limits on the mass-loss rate of $<$10$^{-7}$ $M_{\odot}$yr$^{-1}$. \cite{Welch88} and \cite{Deasy88} conclude that mass-loss during the Cepheid phase is insufficent to explain the observed mass discrepancy. In the case of a 5$M_{\odot}$ Cepheid, mass loss can account for the mass discrepancy if either the lifetime in the IS is ten times longer than derived by \cite{Bono00a} or mass loss is thirty times greater than found by \cite{Deasy88}. Mass loss as an explanation for the mass discrepancy in a 12$M_{\odot}$ Cepheid is more challenging. The mass loss found by Deasy over the entire lifetime of the star can not account for the mass discrepancy and a 600-fold increase in the mass loss rate would be required. 

Recently, \cite{Merand07} found using near-infrared interferometry that, from a sample of four Cepheids all show the presence of some circumstellar material. $\alpha$ Persei, a non-variable supergiant residing in the instability strip, does not show evidence for circumstellar material, suggesting that pulsation does have a role in enhancing mass-loss. The conversion from historical and ongoing mass loss that lead to the circumstellar material and the rate of mass loss is beyond the scope of the study of \cite{Merand07} however. On the current evidence I conclude that mass loss does not present a solution to the Cepheid mass discrepancy.

\subsection{Core He Mass and Internal Mixing}

Cepheid luminosity is critically dependent on the He core mass. The mass of the He core is determined by the extent of the convective core during core H burning. Classical models define the limit to convection via the Schwarzschild criterion. This places the boundary to convection at the radius at which the buoyant force acting on a hot clump of material rising from the convective core drops to zero. However, the temperature and density regime in the vicinity of the convective boundary of the main sequence Cepheid progenitor are such that restorative forces in the region formally stable to convection are mild, and some significant level of overshoot of the classical boundary is expected \citep{Zahn91,Deng07}.

The description of convection is the weakest point in our understanding of the physics of intermediate mass -- massive stars. Numerical modeling of core convection requires a description of the turbulence field at all scales. Three-dimensional hydrodynamical calculations capable of adequate resolution have only recently become a possibility and are in their infancy \citep{Meakin07, Dearborn06, Eggleton03, Eggleton07}. In the absence of a general theory of CCO, which would enable us to calculate the amount of core overshoot for a star of a given mass and chemical composition, a semiempirical phenomenological approach must be used in calculations of stellar evolution. Several computational schemes of various degrees of sophistication in treating the physics of overshoot have been discussed in the literature \citep{Maeder88, Bertelli90, Girardi00, Demarque07, Straka05}. The most common parameterization of core overshoot utilizes the mixing-length formulation, where gas packets progress a distance of $\Lambda_{c}$\footnote{We quantify core overshoot using the formalism of \citet{Bressan81}. Note that $\Lambda_{c}$ is a factor of 2.0 times the overshoot parameter $d_{over}/H_{p}$ in the formalism of the Geneva group \citep{Schaller92, Demarque04}} pressure scale heights into the classically stable region. $\Lambda_{c}$ offers a convenient way to parametrize the extension of the convective core it does not constrain its physical origin. In addition to the CCO mechanism outlined above, rotationally induced mixing can similarly be invoked to bring about a similar range of internal mixing. As shown by \citet{Heger00} and \citet{Meynet00} mixing in the sheer layer formed at the interface between the convective and radiative regions can lead to larger He core masses for massive stars. 

At present we must rely on observation for constraint of $\Lambda_{c}$.  Observational determinations have focussed on populations of intermediate-mass stars (M=5-12$M_{\odot}$), where the signature of CCO is expected to be most clearly seen. The studies of \citet{Mermilliod86} and \citet{Chiosi92} of Galactic open clusters converge on the necessity for mild core overshoot; $\Lambda_{c} \approx 0.5$. A number of subsequent studies have presented evidence both for \citep{Barmina02} and against \citep{Testa99} core overshoot. From studies of the young populous cluster systems of the Magellanic Clouds, \citet{Keller01} found evidence for $\Lambda_{c}=0.62\pm0.11$, while \citet{Cordier02} examined the Magellanic Cloud field population and found the necessity for a level of overshoot between 0.2 and 0.8 pressure scale heights. Measurements with the lowest associated uncertainties are derived from the pulsation modelling of Cepheid light curves by \citet{Keller02, Keller06, Bono02}. \citet{Keller06} finds evidence for a weak dependence of $\Lambda_{c}$ with metallicity with $\Lambda_{c}$ rising from 0.688$\pm$0.009 in the LMC (metallicity 2/5th Solar) to 0.746$\pm$0.009 in the SMC (metallicity 1/5th Solar). 

While radiative opacity and mass-loss have proven insufficient, increased internal mixing remains as the most likely cause of the Cepheid mass discrepancy. The excess of ($17\pm0.05$)\% in evolutionary mass compared to the pulsation mass equates to a uniform level of CCO of $\Lambda_{c}=0.67\pm0.17$. This degree of overshoot is within the range of previous studies discussed above. The scenario I have outlined offers a straightforward explanation for the findings of \citet{Caputo05}, one that uses canonical mass-loss and mild convective core overshoot.

\section{Conclusions}
In this paper I revisit the conclusions of \citet{Caputo05} regarding the cause of the observed discrepancy between the evolution and pulsation masses for Cepheids. \citet{Caputo05} find that Cepheids of 5$M_{\odot}$ have 20\% less mass, as determined from pulsation analysis, than expected from evolutionary calculations, while the evolution and pulsation masses for Cepheids of 14$M_{\odot}$ agree. In order to explain this finding Caputo et al.\ propose a scenario of non-standard mass-loss:- one that sees increased mass-loss at lower masses and drops to negligible mass-loss at $M \ge 14M_{\odot}$. This scheme of mass-loss would be counter to the observationally grounded evidence that accumulated mass-loss by the epoch of core-He burning increases with increasing stellar mass. Furthermore, Caputo et al.\ claim that convective core overshoot is unable to provide a solution as it can not account for the trend of mass discrepancy with Cepheid mass.

The findings of Caputo et al.\ are based on a Cepheid mass-luminosity relationship that proves erroneous when extrapolated to higher masses. In this paper I show that including a full description of the mass-luminosity relation results in a mass discrepancy in which evolutionary masses are ($17\pm5$)\% greater than pulsation masses. The trend of mass discrepancy with Cepheid mass is removed.

I propose that additional internal mixing, as parametrised in the convective core overshooting paradigm, is the primary mechanism giving rise to the mass discrepancy. The level of convective core overshoot so derived is $\Lambda_{c} = 0.67\pm0.17$, a value which agrees with previous determinations from a range of techniques.

\acknowledgments
I thank A. Bressan et al.\ for providing us with unpublished evolutionary models for $\Lambda_{c}=1.0$. I would also like to thank Peter Wood for discussions during the preparation of this paper and Guiseppe Bono for his comments on a draft of this paper.

\bibliography{Keller_ms}

\bibliographystyle{apj}

\clearpage

\clearpage

\clearpage

\end{document}